\documentclass{article}
\usepackage{graphicx}  % standard LaTeX graphics tool;
\usepackage{amsmath}   % For getting proper math eqns
\usepackage[compress]{cite}
\usepackage{amssymb}   % Used for getting various symbols eg. gtrsim, lesssim etc.
\usepackage{bm} % For bold math (esp of lower greek letters)
\usepackage{dcolumn}% Align table columns on decimal point
\usepackage{color}
\usepackage{mathrsfs}
\usepackage{amsfonts}
\usepackage{varioref}
\usepackage{color}
\RequirePackage[colorlinks,citecolor=blue,urlcolor=magenta,linkcolor=blue]{hyperref}
\def\gr{general relativity}
\def\RN{Reissner-Nordstr\"{o}m }
\def\cch{cosmic censorship hypothesis}

\labelformat{section}{Section #1} 
\labelformat{subsection}{Section #1} 
\labelformat{subsubsection}{Section #1}
\labelformat{subsubsubsection}{Section #1}
\labelformat{equation}{Eq.~(#1)} \labelformat{figure}{Fig.~#1} 
\labelformat{subfigure}{Fig.~\thefigure#1} 
\labelformat{table}{Tab.~#1} 
\labelformat{appendix}{Appendix #1}
\title{\bf Overcharging a multi black hole system and cosmic censorship}
\author{Akash K Mishra \footnote{akash.mishra@iitgn.ac.in }$~^{1}$ and Sudipta Sarkar\footnote{sudiptas@iitgn.ac.in}$~^{1}$\\
\small{$~^{1}$ Indian Institute of Technology, Gandhinagar-382355, Gujarat, India}\\
}
\date{ }  %% This command  will supress printing the date. 
\begin{document}
  
\maketitle
%%%%%%%%%%%%%%%%%%%%%%%%%%%%%%%%%%%%%%%%%%%%%%%%%%%%%%%%%%%%%%%%%%%%%%%%%%%%%%%%%%%%%%%%%%%%%%%%%%%
%%%%%%%%%%%%%%%%%%%%%%%%%%%%%%%%%%%%%%%%%%%%%%%%%%%%%%%%%%%%%%%%%%%%%%%%%%%%%%%%%%%%%%%%%%%%%%%%%%%
%%%%%%%%%%%%%%%%%%%%%%%%%%%%%%%%%%%%%%%%%%%%%%%%%%%%%%%%%%%%%%%%%%%%%%%%%%%%%%%%%%%%%%%%%%%%%%%%%%%
\begin{abstract}
We study the generalization of the gadenken experiment of overcharging an extremal black hole proposed by Wald in the context of a multi black hole solution. In particular, we attempt to overcharge a system of two extremal black holes via test particle absorption to produce a system involving a black hole and a naked singularity. If such a process is possible, then this would be a potential violation of the \cch. However, we find that, analogous to Wald's result for a single charged black hole, such a test particle which can expose the singularity, would not be able to enter the horizon. This provides an interesting and non-trivial example that supports the validity of the \cch in four dimensional general relativity.

\end{abstract}
%%%%%%%%%%%%%%%%%%%%%%%%%%%%%%%%%%%%%%%%%%%%%%%%%%%%%%%%%%%%%%%%%%%%%%%%%%%%%%%%%%%%%%%%%%%%%%%%%%%
%%%%%%%%%%%%%%%%%%%%%%%%%%%%%%%%%%%%%%%%%%%%%%%%%%%%%%%%%%%%%%%%%%%%%%%%%%%%%%%%%%%%%%%%%%%%%%%%%%%
%%%%%%%%%%%%%%%%%%%%%%%%%%%%%%%%%%%%%%%%%%%%%%%%%%%%%%%%%%%%%%%%%%%%%%%%%%%%%%%%%%%%%%%%%%%%%%%%%%%
%\newpage
%%%%%%%%%%%%%%%%%%%%%%%%%%%%%%%%%%%%%%%%%%%%%%%%%%%%%%%%%%%%%%%%%%%%%%%%%%%%%%%%%%%%%%%%%%%%%%%%%%%
%%%%%%%%%%%%%%%%%%%%%%%%%%%%%%%%%%%%%%%%%%%%%%%%%%%%%%%%%%%%%%%%%%%%%%%%%%%%%%%%%%%%%%%%%%%%%%%%%%%
%%%%%%%%%%%%%%%%%%%%%%%%%%%%%%%%%%%%%%%%%%%%%%%%%%%%%%%%%%%%%%%%%%%%%%%%%%%%%%%%%%%%%%%%%%%%%%%%%%%
\section*{Introduction}

Gravity is best described in the framework of general relativity as the manifestation of the curvature of space and time. Even after more than one hundred years, \gr\ still stands as one of the most beautiful theory human mind has ever produced and has often surprised us by it's various bold yet interesting predictions like black holes, gravitational waves, expansion of the universe, etc. General relativity has passed all the observational tests covering
a wide range of scales starting from solar system measurements like the perihelion precession of mercury, up to tests at the cosmological level. Despite having such theoretical beauty and observational consistencies, \gr\ certainly cannot be considered as the final theory of gravity because of several limitations. One among such limitations of \gr\ is the inevitable occurrence of singularities in the solutions of Einstein's equation\cite{Wald:1984rg}. Gravitational singularities are well known to physicists ever since the discovery of the Schwarzschild black hole solution and has been one of the major themes of research. All the physically relevant quantities diverge at the singularity, and the underlying physical theory loses its predictive power. In fact, since the notion of space and time becomes ill-defined at the singularity, the definition of singularities in general relativity is far from obvious. Gravitational Singularities as of now, are best understood in terms of the geodesic incompleteness of causal curves \cite{Wald:1984rg,hawking_ellis_1973}. 
\\

The issue of occurrence of singularities in generic solutions of Einstein's equation has been pursued by numerous researchers over many decades, and the most definitive work in this regard is the Hawking and Penrose singularity theorems \cite{Wald:1984rg,hawking_ellis_1973}. These theorems are a set of results which asserts that the solution of Einstein's equation with physically reasonable initial data, would always evolve to a singularity via gravitational collapse. One of the major key insight in classical \gr\ is the realization that such a singularity which results from the gravitational collapse of an object may always be hidden inside the event horizon of a black hole and not accessible to an outside observer. This is known as the weak cosmic censorship conjecture (WCCC), first stated by Penrose\cite{Penrose:1969pc,hawking_ellis_1973} and is essential in ensuring the predictability of physical laws outside the black hole. More precisely, the WCCC is the statement that regular initial data cannot be evolved by Einstein's equation to a naked singularity solution; a spacetime singularity without an event horizon. For some interesting reviews on the cosmic censorship conjecture and gravitational collapse, refer to \cite{Wald:1997wa,doi:10.1142/S0217732302007570,Clarke_1994} and Refs. therein. 
\\

Considering the complexity involved in Einstein's equations, it is challenging and almost impossible to come up with a general proof of the statement, and therefore cosmic censorship as of now remain a major open problem in classical \gr. However, attempts have been made by numerous researchers to approach the problem in the other way around, i.e., by looking for possible counterexample of the hypothesis. The strategy is to start with a black hole solution as an initial configuration (extremal or non-extremal) and look for any possible physical process that would give rise to a naked singularity in the final state. Such an analysis was pioneered by Wald in his seminal paper \cite{Wald}, where he attempts to overcharge and over-spun an extremal Kerr-Newman black hole via the process of charged test particle absorption to produce a naked singularity. Interestingly, the electromagnetic repulsion between the test particle and the black hole turns out to be enough to prevent the black hole from capturing such a particle. 
\\

The WCCC is supported as well as challenged by several such examples and counterexamples. Extending Wald's calculation further, an analogous result of overcharging a Kerr-Newman black hole in the presence of both electric and magnetic charged was studied in \cite{Semiz:1990fm}. Overcharging via test field absorption instead of test particle was pursued in \cite{Semiz:2005gs,Toth:2011ab,Nat_rio_2016}. Various other attempts of overcharging have been carried out in last few decades over a wide range of black hole solutions in \gr\ in order to look for any possible violation of cosmic censorship \cite{Hod:1999kn,Hod:2008zza,Chirco:2010rq,BouhmadiLopez:2010vc,Saa:2011wq,Rocha:2014jma,Fairoos:2017lnm,Revelar:2017sem,An:2017phb,Yu:2018eqq,Jana:2018knq,Shaymatov:2018fmp,Ge:2017vun}. \\

In Ref \cite{Hubeny:1998ga}, Hubeny showed that starting with a slightly non-extremal charged black hole, it is possible to create a naked singularity in the final configuration via charged particle absorption. Such near-extremal black hole overcharging problem was further pursued in\cite{Hod:2002pm,Matsas:2007bj,Jacobson:2009kt}. Further such counterexamples to the WCCC can be found in \cite{Lehner:2010pn,Gao:2012ca}. 
However, it is also argued that the violation of the WCCC can be avoided by taking into account the back-reaction effect of the test particle. For follow-up works on the WCCC with the self force effect refer to \cite{Barausse:2010ka,Barausse:2011vx,Isoyama:2011ea, Zimmerman:2012zu, Colleoni:2015afa}.\\

Recently, in \cite{Sorce:2017dst}, the authors have presented a general analysis of the overcharging problem for arbitrary matter falling into the black hole (extremal and non-extremal) by taking into account up to second order variation of the parameters of the black hole. The main result of \cite{Sorce:2017dst} is that overcharging of extremal as well as non-extremal black hole is not possible, as long as the matter stress-energy tensor satisfies the Null Energy Condition.
\\

Hence, considering the current status and in the absence of a general proof, it is very much desirable for one to look for various other possible tests of the \cch. Such tests as of now available in the literature only involve the overcharging of a single black hole configuration, and till now, no attempts have been made to study the WCCC in the context of a system comprising more than a single black hole. This motivates us to test the \cch\ in a multi black hole setting. In this article we attempt to study the problem of overcharging via test particle absorption in the context of the well known Majumdar-Papapatrou black hole solution, which corresponds to a system of two extremal black holes in equilibrium\cite{PhysRev.72.390,10.2307/20488481,hartle1972,Semerak:2016gfz}. The final configuration that we would be interested in is a di-hole, which is a stable system of a non-extremal black hole and a naked singularity and exists as a non-trivial solution of the Einstein-Maxwell's equations \cite{Alekseev:2007gt}. Essentially, we would attempt to address the following question: Is it possible to obtain a stable di-hole configuration in the final state, starting from two extremal black holes in equilibrium? The question of stability is far richer in a di-hole configuration than that of a single black hole and requires additional condition on the parameters involved i.e., the mass($m_i$), charge($q_i$) and distance between the two sources\cite{Alekseev:2007gt}. Also, the spacetime structure in such a system is vastly different than that of a single black hole because of different symmetry of the spacetime\cite{Ryzner:2015jda}. A priori there is no reason to believe that the WCCC would hold for such cases and requires careful analysis.
\\

To our surprise, even for such a nontrivial system, the overcharging is found to be impossible and does not lead to any violation of the \cch. Specifically, we show that it is impossible to obtain an equilibrium di-hole system in the final configuration via test particle absorption. We also show that one cannot even overcharge to obtain a naked singularity without using the equilibrium condition. The result seems to be a strong and nontrivial example in favor of the \cch. In this analysis, we neglect the self-force effect.

\section*{Overcharging a Majumdar-Papapatrou configuration}

The equilibrium configuration of multiple massive objects in Newtonian physics is vastly different than the same situation in \gr\cite{Bonnor:1981ag,Ohta:1982dq,Tomimatsu:1983ph,Azuma:1994nd,Bini:2006pk,Manko:2007hi}. The only equilibrium configuration in \gr\ is known to be either a system of two extremal black holes or a combination of a nonextremal black hole and a naked singularity satisfying some stability conditions. The former one is the well known Majumdar-Papapatrou solution and the later we would refer as a di-hole\cite{Alekseev:2007gt}. As of now, the di-hole configuration represents the only solution of the Einstein-Maxwell equation known in literature consisting of two non-extreme sources in equilibrium. We start this section by providing an overview of the di-hole geometry and subsequently present an analysis of the motion of test particles on such background. In terms of Wyel cylindrical coordinate, the line element can be written as\cite{Alekseev:2007gt},

%%%%%%%%%%%%%%%%%%%%%%%%%%%%%%%%%%%%%%%%
 \begin{eqnarray}
ds^2=-H(\rho,z)dt^2+f(\rho,z)(d\rho^2+dz^2)+\frac{\rho^2}{H}d\phi^2 \\
A_t=\Phi(\rho,z),~~~~A_\rho=A_z=A_\phi=0
\end{eqnarray}
%%%%%%%%%%%%%%%%%%%%%%%%%%%%%%%%%%%%%%%%
Where, $A_\mu$ is the electromagnetic vector potential, and $z$ is the symmetry axis. Considering the symmetry of the problem, it is convenient to express the metric components in terms of bipolar coordinates $(r_1,\theta_1)$ and $(r_2,\theta_2)$, centered at both the sources respectively and spanning the entire spacetime, which can be achieved by the following coordinate transformations,

%%%%%%%%%%%%%%%%%%%%%%%%%%%%%%%%%%%%%%%%
\begin{eqnarray}
\rho=\sqrt{(r_1-m_1)^2-\sigma_1^2}\sin{\theta_1},~~~~~z=z_1+(r_1-m_1)\cos{\theta_1}\\
\rho=\sqrt{(r_2-m_2)^2-\sigma_1^2}\sin{\theta_2},~~~~~z=z_2+(r_2-m_2)\cos{\theta_2}
\end{eqnarray}
%%%%%%%%%%%%%%%%%%%%%%%%%%%%%%%%%%%%%%%%
The notations in the above expression are,

%%%%%%%%%%%%%%%%%%%%%%%%%%%%%%%%%%%%%%%%
\begin{eqnarray}
&&\sigma_1^2=m_1^2-e_1^2+2e_1\gamma ,~~~~~~~~~~~\sigma_2^2=m_2^2-e_2^2-2e_2\gamma\\
&&\gamma=\frac{m_2e_1-m_1e_2}{l+m_1+m_2},~~~~~~~~~~~~~~~~~~l=z_2-z_1\label{gamma}
\end{eqnarray}

%%%%%%%%%%%%%%%%%%%%%%%%%%%%%%%%%%%%%%%%
Here $m_1,m_2$ and $e_1,e_2$ represents the physical mass and charge of the two non-extreme \RN\ sources located at points $z_1$ and $z_2$ respectively on the symmetry axis, with `$l$' being the distance between them. For this configuration, it turns out that the region between the two sources contains conical singularities (struts) at each point on the symmetry axis. However, the condition,
%%%%%%%%%%%%%%%%%%%%%%%%%%%%%%%%%%%%%%%%
\begin{equation}
m_1m_2=(e_1-\gamma)(e_2+\gamma)\label{eqb}
\end{equation}
%%%%%%%%%%%%%%%%%%%%%%%%%%%%%%%%%%%%%%%%
ensures the equilibrium between the two sources without the presence of any such struts on the symmetry axis. Note that, when one of the sources is absent, the equilibrium condition \ref{eqb} is trivially satisfied. \\
The components of metric ($H,f$) and electrostatic potential ($\Phi$) in this bi-polar coordinate system reads\cite{Alekseev:2007gt},
%%%%%%%%%%%%%%%%%%%%%%%%%%%%%%%%%%%%%%%%
\begin{eqnarray}
H&=&\frac{\left[(r_1-m_1)^2-\sigma_1^2+\gamma^2\sin^2{\theta_2}\right]\times \left[(r_2-m_2)^2-\sigma_2^2+\gamma^2\sin^2{\theta_1}\right]}{[r_1r_2-(e_1-\gamma-\gamma\sin{\theta_1})(e_2+\gamma-\gamma\sin{\theta_2})]^2}\\
\nonumber\\
\Phi &=&-\frac{\left[(e_1-\gamma)(r_2-m_2)+(e_2+\gamma)(r_1-m_1)+\gamma(m_1\cos{\theta_1}+m_2\cos{\theta_2})\right]}{[r_1r_2-(e_1-\gamma-\gamma\sin{\theta_1})(e_2+\gamma-\gamma\sin{\theta_2})]}\\
\nonumber\\
f&=&\frac{[r_1r_2-(e_1-\gamma-\gamma\sin{\theta_1})(e_2+\gamma-\gamma\sin{\theta_2})]^2}{[(r_1-m_1)^2-\sigma_1^2\cos^2{\theta_1}]\times [(r_2-m_2)^2-\sigma_2^2\cos^2{\theta_2}]}
\end{eqnarray}
%%%%%%%%%%%%%%%%%%%%%%%%%%%%%%%%%%%%%%%%
Note that $\sigma_k^2=0$ ($k=1,2$ for both sources) represents the extremal limit and the above line element reduces to that of a Majumdar-Papapatrou solution consisting of two extremal black holes. However, $\sigma_{k}^2>0$ represents a non-extreme black hole whereas $\sigma_{k}^2<0$ corresponds to a naked singularity. The location of the horizon can be obtained by setting $H=0$, which corresponds to $r_1 = m_1$ and $r_2=m_2$ in the extremal limit or equivalently $\rho=0$. \\

Having discussed the geometry of a di-hole system, now we would like to understand the overcharging problem and the cosmic censorship in this context. To that end, let us start with a system of two extremal black holes ($\sigma_{1(i)}^2=\sigma_{2(i)}^2 = 0,\gamma =0$) as our initial configuration. Let the first extremal black hole absorb a test particle of mass $\delta m_1$, charge $\delta e_1$ and turn into a non-extreme black hole of final mass $m_1+\delta m_1$ and final charge $e_1+\delta e_1$. The test particle assumption is ensured by the conditions $\delta m_k<<m_k$ and $\delta e_k<<e_k$. For a black hole in the final state, we have $\sigma_{1(f)}^2>0$, which further reduces to,

%%%%%%%%%%%%%%%%%%%%%%%%%%%%%%%%%%%%%%%%
\begin{eqnarray}
(\delta m_1 - \delta e_1)\,>\, -\delta \gamma\label{bh_cond_1}
\end{eqnarray} 
%%%%%%%%%%%%%%%%%%%%%%%%%%%%%%%%%%%%%%%%
And the second extremal black hole absorb another test particle of mass $\delta m_2$, charge $\delta e_2$ and turn into a naked singularity of final mass $m_2+\delta m_2$ and final charge $e_2+\delta e_2$. For a naked singularity in the final state we have $\sigma_{2(f)}^2<0$, which leads to,
%%%%%%%%%%%%%%%%%%%%%%%%%%%%%%%%%%%%%%%%
\begin{eqnarray}
(\delta m_2 - \delta e_2)\,<\, \delta \gamma\label{ns_cond_1}
\end{eqnarray} 
%%%%%%%%%%%%%%%%%%%%%%%%%%%%%%%%

\ref{bh_cond_1} and \ref{ns_cond_1} represents the constraints on the mass and charge of both the test particles for which the final state would correspond to a configuration of a non-extreme black hole and a naked singularity, which can be regarded as a potential violation of the \cch. 
Now it is important to understand if the initial extremal black holes can capture such test particles satisfying the above two condition. We carry out this analysis by studying the geodesic motion of test particles on the Majumdar-Papapetrou background.
Let us start with the motion of first test particle with mass $\delta m_1$ and charge $\delta e_1$. For a time-like geodesic, with metric signature $diag(-,+,+,+)$ we have,
\begin{eqnarray}
-1=-H\dot{t}^2+f(\dot{\rho}^2+\dot{z}^2)+\frac{\rho^2}{H}\dot{\phi}^2\label{timelike}
\end{eqnarray}
where the derivatives are w.r.t. the proper time $\tau$ of the test particle. The energy of the particle is given by,
\begin{eqnarray}
-\delta m_1=\delta e_1\Phi - \dot {t}\,H
\end{eqnarray}
Now we eliminate $\dot{t}$ in \ref{timelike}, which reduces to,
\begin{eqnarray}
(\delta m_1+\delta e_1\Phi)^2-H-\rho^2\dot{\phi}^2=Hf(\dot{\rho}^2+\dot{z}^2)
\end{eqnarray}
Since our initial configuration is a system of two extremal source we can use $HF=1$ to simplify the above expression further. Writing $\dot{\rho}$ and $\dot{z}$ in terms of $\dot{r_1}$ and $\dot{\theta_1}$,
\begin{eqnarray}
\dot{\rho}&=&\sqrt{(r_1-m_1)^2-\sigma_1^2}\cos{\theta_1}\dot{\theta_1}+\frac{(r_1-m_1)\dot{r_1}}{\sqrt{(r_1-m_1)^2-\sigma_1^2}}\sin{\theta_1}\\
\dot{z}&=&\dot{r_1}\cos{\theta_1}-(r_1-m_1)\sin{\theta_1}\dot{\theta_1}
\end{eqnarray}
%%%%%%%%%%%%%%%%%%%%%%%%%%%%%%%%%%%%%%%%
At the extremal horizon($r_1 = m_1$) of the first black hole, one has $\Phi = -1$ and $H =0$ and the geodesic equation further reduces to,
\begin{eqnarray}
(\delta m_1-\delta e_1)^2=\dot{r_1}^2
\end{eqnarray}
%%%%%%%%%%%%%%%%%%%%%%%%%%%%%%%%%%%%%%%%
Now, in order to cross the extremal horizon, the test particle must have $\dot{r}_1<0$ at the horizon. Further note that, the energy of the particle measured at infinity is positive, where the potential vanishes, and we have 

\begin{eqnarray}
\dot{r_1}=-(\delta m_1-\delta e_1)
\end{eqnarray}
Therefore the mass and charge of the test particle must satisfy $\delta m_1>\delta e_1$ in order to get captured by the black hole. Similarly, the second test particle must satisfy $\delta m_2>\delta e_2$ in order to cross the horizon of the second black hole. Also notice that, so far we have not used the equilibrium condition to hold in the final configuration and we shall study both cases of presence and absence of struts on the symmetry axis separately. Let us start with the case, where the equilibrium condition holds in the final configuration, i.e., $m_{1(f)}m_{2(f)}\, =\, (e_{1(f)}-\gamma_f)(e_{2(f)}+\gamma_f)$, which ensures absence of any struts on the symmetry axis between the two sources and using the fact that $\gamma =0$ for the initial configuration it further reduces to,
\begin{equation}
\gamma_f =\delta\gamma = \frac{m_1(\delta m_2 - \delta e_2) + m_2(\delta m_1 - \delta e_1)}{(m_1-m_2)}\label{stability}
\end{equation}
%%%%%%%%%%%%%%%%%%%%%%%%%%%%%%%%%%%%%%%
However, from the general expression in \ref{gamma}, by taking the first order variation of $\gamma$, we obtain, 
%%%%%%%%%%%%%%%%%%%%%%%%%%%%%%%%%%%%%%%
\begin{equation}
 \delta\gamma = \frac{m_1(\delta m_2 - \delta e_2) - m_2(\delta m_1 - \delta e_1)}{(l+m_1+m_2)}\label{general_gamma}
\end{equation}
%%%%%%%%%%%%%%%%%%%%%%%%%%%%%%%%%%%%%%%
Now, assuming the equilibrium condition to hold in the final configuration, we equate \ref{stability} and \ref{general_gamma} to solve for `$l$', in terms of the variations, which gives rise to,
\begin{equation}
\delta_1 + \delta_2 = -\frac{l(m_1\delta_2+m_2\delta_1)}{2 m_1 m_2}\label{variation}
\end{equation}
%%%%%%%%%%%%%%%%%%%%%%%%%%%%%%%%%%%%%%%%%%%%%%%%%%%%%%%%%%%%%%%%%%%%%%%
where, $\delta_1 = \delta m_1 - \delta e_1$ and $\delta_2 = \delta m_2 - \delta e_2$. It is evident from the above expression that, as far as the test particles are entering the horizons, i.e., $\delta_1>0$ and $\delta_2>0$, \ref{variation} can not be satisfied for the positive value of the length $l$ and hence the equilibrium condition will not hold in the final configuration. Therefore, although we have started with a Majumdar-Papapetrou configuration initially, it would not be possible to maintain the equilibrium condition in the final configuration by test particle absorption. This rules out the possibilities of obtaining any equilibrium system (two extremal black holes, a black hole \& a naked singularity or two naked singularities) by starting from a Majumdar-Papapetrou system. Note that, here the equilibrium condition ensures the absence of any struts on the symmetry axis.\\

%%%%%%%%%%%%%%%%%%%%%%%%%%%%%%%%%%%%%%%%%%%%%%%%%%%%%%%%%%%%%%%%%%%%%%%
Having illustrated that, test particle absorption would not lead to any equilibrium configuration; now we would like to understand, whether one can produce any naked singularity in the final configuration without demanding the equilibrium condition. In other words, we would like to see if we can produce a black hole and a naked singularity in the final configuration, with struts on the symmetry axis. To that end, we substitute the general first order variation of $`\gamma$'(which doesn't use the equilibrium condition) from \ref{general_gamma} in \ref{bh_cond_1} and \ref{ns_cond_1} to obtain,%%%%%%%%%%%%%%%%%%%%%%%%%%%%%%%%%%%%%%%%%%%%%%%%%%%%%%%%%%%%%%%%%%%%%%%
\begin{equation}
(\delta m_1 - \delta e_1)> -\left(\frac{m_1}{l+m_1}\right)(\delta m_2 - \delta e_2)\label{bh_cond_2}
\end{equation}

\begin{equation}
(\delta m_2 - \delta e_2)< -\left(\frac{m_2}{l+m_2}\right)(\delta m_1 - \delta e_1)\label{ns_cond_2}
\end{equation}
%%%%%%%%%%%%%%%%%%%%%%%%%%%%%%%%%%%%%%%%%%%%%%%%%%%%%%%%%%%%%%%%%%%%%%%
Here, \ref{bh_cond_2} and \ref{ns_cond_2} represents the conditions for obtaining a non-extremal black hole and a naked singularity in the final configuration respectively. Note that, the condition of source `1' becoming a black hole explicitly depends on the constraint of test particle entering the second source and vice-versa. 
It is not difficult to realize that, the condition for obtaining a non-extremal black hole is in consistent with the condition for the test particle entering the horizon. However, the condition for obtaining a naked singularity is not compatible with the test particle entering the horizon. Therefore, it is not possible to obtain any naked singularity, with or without using the equilibrium condition via test particle absorption by starting from two extremal sources. Hence, no violation of the WCCC can be achieved in the context of a Majumdar-Papapetrou configuration.

\section*{Discussions}
Even after more than five decades of its proposal by Penrose, the \cch\ still remain one unsolved puzzle in classical \gr. One of the potential counterexample to the hypothesis is the overcharging of a slightly non-extremal black hole to produce naked singularity. However, it is widely believed that such processes would not occur when the backreaction effect of the test particle is taken into account. Although the conjecture lacks a general proof, there seems to be growing number of evidence in its favor. The general theme of approach in this regard has been to look for any possible counterexample to the hypothesis, which may results from regular initial data via some physical process. Following the gadenken experiment of Wald to overcharge an extremal Kerr-Newman black hole via test particle absorption, in this article, we attempted to generalize this result to a multi black hole solution of general relativity. 
In particular, we studied the overcharging problem of a system of two extremal black holes represented by a Majumdar-Papapetrou solution in the initial configuration to produce a stable system of two non-extreme sources consisting of a black hole and a naked singularity in the final configuration. If such a physical process happens to exists, this will correspond to a potential violation of the \cch. Because of distinct geodesic structure than that of a single source and an additional equilibrium condition, it certainly not evident whether the \cch\ would still hold for such settings. Surprisingly, we found that such test particles which would expose the naked singularity upon absorption by the black hole, would not be able to enter the black hole horizon and the \cch\, even in such non-trivial setting remains intact. We have also shown that test particle absorption would not lead to any equilibrium final configuration (absence of struts). This is an important result supporting the validity of cosmic censorship since the di-hole system is the only equilibrium solution of general relativity involving two non-extreme sources known in the literature.  An interesting direction to pursue further would be to carry out this analysis for a system of $N$-extremal charged black holes. 

\section*{Acknowlwdgement}
We thank Avirup Ghosh for many helpful discussions. Research of SS is supported by the Department of Science and Technology, Government of India under the SERB Fast Track Scheme for Young Scientists (YSS/2015/001346). 

\bibliography{WCCC}

\bibliographystyle{./utphys1}

\end{document}